\documentclass{jnmp}

\usepackage{amsmath}
\usepackage{graphicx}

\JNMPnumberwithin{equation}{section}

\def\la{\langle}
\def\ra{\rangle}
\def\C{\mathsf{C}}
\def\D{\mathsf{D}}
\def\E{\mathsf{E}}

\def\1{\mathsf{1}}
\def\({\left(}
\def\){\right)}
\def\Z{\mathbb{Z}}
\def\T{\mathsf{T}}

\begin{document}

\renewcommand{\evenhead}{M Uchiyama and M Wadati}
\renewcommand{\oddhead}{Correlation function of ASEP with open boundaries}

\thispagestyle{plain}

\Name{Correlation Function of Asymmetric Simple Exclusion Process with Open Boundaries}

\label{firstpage}

\Author{\Large{Masaru UCHIYAMA and Miki WADATI}}

\Address{Department of Physics, Graduate School of Science, University of Tokyo, \\
Hongo 7-3-1, Bunkyo-ku, Tokyo, 113-0033, Japan\\
E-mail: uchiyama@monet.phys.s.u-tokyo.ac.jp, wadati@monet.phys.s.u-tokyo.ac.jp
}


\hspace{1cm}
\begin{center}
{\it Dedicated to Professor Francesco Calogero on the occasion of \\
his seventieth birthday.}
\end{center}
\hspace{1cm}

\begin{abstract}
\noindent
We investigate the correlation functions of the one-dimensional Asymmetric Simple Exclusion Process (ASEP) with open boundaries. The conditions for the boundaries are made most general. The correlation function is expressed in a multifold integral whose behavior we study in detail. We present a phase diagram of the correlation length. For the case the correlation length diverges, we further give the leading terms of the finite-size correction. 
\end{abstract}

\section{Introduction}

As we all appreciate, exactly solvable models have been playing important roles for understanding nonlinear physics in various areas. 
They make it possible to calculate physical quantities exactly so that we can enjoy arguments beyond approximation. 
What is more, the integrability of models often serves a deeper mathematical structure. 
For one thing, we consider the one-dimensional Asymmetric Simple Exclusion Process (ASEP). The ASEP is an exactly solvable model of stochastic process regarded as a primitive model of kinetics of biopolymerization, traffic flow, and so on. 
The model is known to have nonequilibrium properties in the stationary situation and is studied extensively for the nonequilibrium statistical mechanics 
\cite{Ligg,Ligg2,Privman,SZ,Schuetz,Sp,S-W}. 
In the mathematical aspect, it has a connection with the theory of the $q$-orthogonal polynomial \cite{SS99,BECE,USW04}. 

The nonequilibrium statistical mechanics has attracted a renewed interest among physicists these days. 
Numerical experiments are more available than before and a lot of interesting applications are found out. 
Despite of these developments, the central principle of the nonequilibrium statistical mechanics is still unclear. 
In order to unveil the central principle, we need to look for basic concepts and clarify mathematical structures. 
There are some attempts ongoing for this end such as a construction of thermodynamics for the steady state which is far from equilibrium \cite{SST}. 
At the same time, we may have recourse to reliable exactly solvable models and analyse the behaviors in detail. 
On the way, we can define quantities which are essential for the general theory. 
For example, for the ASEP we find a definition of the partition function which really is a generating function of the bulk quantities as is familiar in the standard equilibrium statistical mechanics. 
In this sense, exact analytical arguments for individual models are of great value also in the study of the nonequilibrium statistical mechanics. 

Physically and mathematically, boundary conditions are important. 
In the case with open boundaries, an interesting feature of the ASEP is the connection with the Askey-Wilson polynomial. The Askey-Wilson polynomial is a one-variable $q$-orthogonal polynomial. There is a set of lists called the Askey-scheme where many other orthogonal polynomials are recovered by parameter reduction from the Askey-Wilson polynomial \cite{AW,GR,Koekoek}. 
As to physics, the Askey-Wilson polynomial also appears in various integrable models; 
the problem of Bloch electrons in a magnetic field on a lattice called the Azbel-Hofstadter problem which is closely related with the quantum Hall effect \cite{WZ95}, 
the XXZ model with open boundaries \cite{Ismail04}, 
the sine-Gordon model with a boundary and the more general algebraic formulation manipulating them \cite{Bas04-1,Bas04-2}. 
Through the connection with such a mathematical structure, we can obtain more information about the systems than we expect. 

The physics of the ASEP is rich. 
Because of the nonlinearity, the formation of shock profile happens \cite{DJLS}. 
In fact, macroscopically the ASEP obeys the noisy Burgers equation. 
In the setting of an infinite system, the current fluctuation can be calculated based on the Random Matrix Theory through the statistics of the Young diagram \cite{Jo2000,PS2002a}. 
The Random Matrix Theory is a theory of universality classes. 
It is found that the current fluctuation of the ASEP belongs to the KPZ nonequilibrium universality class \cite{KPZ}. 
For a system with open boundaries, by the boundary effect there occurs phase transitions even in the one-dimensional system. 
In contrast, according to the Mermin-Wagner theorem, a phase transition does not take place in a one-dimensional equilibrium system. 
We can give a lot of fascinating phenomena like this.

The aim of this paper is to study the correlation functions for the steady state of the ASEP with open boundaries. The ingredient of our analysis is the integral expression of the $n$-point function. 
The organization of the paper is as follows. 
In section 2, the model and related quantities of the ASEP are introduced. 
In section 3, based on \cite{USW04}, some known results are briefly summarized. It is remarked that the $n$-point function is expressed as a multifold integral. 
In section 4, the correlation functions for the steady state of the ASEP are studied and a phase diagram for the correlation length is presented. 
For the case where the correlation length diverges, we further inquire the integral representing the correlation function and obtain the leading terms of the finite-size correction. 
Section 5 is given for conclusion. 

\newpage

\section{Asymmetric Simple Exclusion Process (ASEP)}

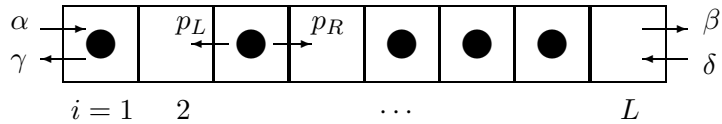
\begin{figure}[htb]
\begin{center}
\setlength{\unitlength}{1mm}
\begin{picture}(100,15)
\multiput(10,0)(10,0){8}{\framebox(10,10)}
\put(7,7){\vector(1,0){6}}
\put(13,3){\vector(-1,0){6}}
\put(87,7){\vector(1,0){6}}
\put(93,3){\vector(-1,0){6}}
\put(35,5){\circle*{4}}
\put(55,5){\circle*{4}}
\put(65,5){\circle*{4}}
\put(15,5){\circle*{4}}
\put(75,5){\circle*{4}}
\put(32,5){\vector(-1,0){5}}
\put(38,5){\vector(1,0){5}}
\put(3,7){$\alpha$}
\put(3,2){$\gamma$}
\put(95,7){$\beta$}
\put(95,1){$\delta$}
\put(25,7){$p_L$}
\put(43,7){$p_R$}
\put(11,-5){$i=1$}
\put(25,-5){$2$}
\put(52,-5){$\cdots$}
\put(84,-5){$L$}
\end{picture}
\end{center}
\caption{The ASEP with open boundaries. 
Black circles and boxes represent particles and lattice sites, respectively.}
\label{asepfig}
\end{figure}

We introduce the model of the one-dimensional Asymmetric Simple Exclusion Process (ASEP). 
Roughly speaking, the model describes randomly hopping 
particles with exclusion interaction on a one-dimensional lattice. 
The term "asymmetric" refers to the asymmetric hopping of particles and "simple exclusion" refers to their hard-core exclusion interaction. 
We consider a finite system of size $L$ with open boundaries at both left and right sides. 
On each site $i=1,\cdots,L$ of the lattice there can be only up to one particle, that is, if we denote by $\tau_i$ the particle number at site $i$, each site is either occupied by one particle $\tau_i=1$ or empty $\tau_i=0$. 
 A particle hops to the nearest right 
(resp. left) site with the probability $p_Rdt$ (resp. $p_Ldt$) 
during an infinitesimal time $dt$ if the target site is empty. 
By rescaling time, we can set $p_R=1$ and $p_L=q$. 
At the boundaries, a particle enters the 
system at the left (resp. right) boundary with rate 
$\alpha$ (resp. $\delta$) if the site $1$ (resp. $L$) is empty, and leaves 
at the left (resp. right) boundary with rate $\gamma$ (resp. $\beta$) 
if the site $1$ (resp. $L$) is occupied. 
The boundary conditions are made most general containing four independent 
parameters $\alpha, \beta, \gamma$ and $\delta$. 
Figure \ref{asepfig} illustrates the setup of the system. 
It is known that the system behaves very differently for the symmetric hopping case $q=1$ and the asymmetric case $q\neq1$. We only consider the asymmetric case in the following. We can restrict $q<1$ because particle-hole exchange simply gives $q>1$. 

The state of this system is labelled by the configuration of particles $\{\tau_i\}=(\tau_1, \tau_2, \cdots ,\tau_L)$ where $\tau_i=1,0$. Then, the time evolution of the probability distribution $P(\{\tau_i\};t)$ is governed by the master equation
\begin{align}
\frac{d}{dt}\vert P(\{\tau_i\};t)\ra = H\vert P(\{\tau_i\};t)\ra .
\end{align}
Here we used the bra-ket notation. 
The time evolution operator $H$ is a matrix whose elements are the transition probability rates, and is defined explicitly by 
\begin{align}
H&=\sum_{i=1}^L \left[q\sigma_i^+\sigma_{i+1}^- 
+\sigma_i^-\sigma_{i+1}^+ 
-\frac{q}{4}(1-\sigma_i^z)(1+\sigma_{i+1}^z)
-\frac{1}{4}(1+\sigma_i^z)(1-\sigma_{i+1}^z)\right] \notag\\
& +\alpha\sigma_1^++\gamma\sigma_1^--\frac{\alpha}{2}(1+\sigma_1^z)
-\frac{\gamma}{2}(1-\sigma_1^z) 
+\delta\sigma_L^++\beta\sigma_L^--\frac{\delta}{2}(1+\sigma_L^z)
-\frac{\beta}{2}(1-\sigma_L^z)
\end{align}
where $\sigma_i^a (a=x,y,z)$ is the Pauli matrix acting on the $i$-th site in the basis $\tau_i=1,0$ and $\sigma_i^\pm=(\sigma_i^x\pm\sqrt{-1}\sigma_i^y)/2$. 
From the general properties of the time evolution operator, there exists a unique steady state: the state which is stationary under the time evolution, or the highest eigenstate of $H$ with eigenvalue zero. 
In this paper, we focus on the steady state. 

As is obvious from the explicit form of the time evolution operator, the one-dimensional ASEP can be regarded as a non-hermitian generalization of spin chain models in which the spin-flip occurs asymmetrically. 
There are a lot of techniques to solve spin chain models. For example the Bethe ansatz method is the representative one. For our purpose, the matrix method is useful \cite{DEHP}. According to the matrix method, the probability distribution of the steady state which we denote by $P(\{\tau_i\})$ is expressed in the matrix product form: 
\begin{align}
\label{eqn:MPA}
P(\{\tau_i\})=
\frac{1}{Z_L}\la W\vert 
\prod_{i=1}^{L\atop\longrightarrow}(\tau_i\D+(1-\tau_i)\E) 
\vert V\ra.
\end{align}
Here $\D$ and $\E$ are matrices; $\D$ (resp. $\E$) 
corresponds to the occupation (resp. emptiness) of a particle. 
$\la W\vert$ and $\vert V\ra$ are vectors; $\la W\vert$ 
(resp. $\vert V\ra$) corresponds to the left (resp. right) boundary. 
One can show that (\ref{eqn:MPA}) gives the exact steady 
state of the system if the matrices and the vectors satisfy
\cite{DEHP} 
\begin{align}
&\D\E-q\E\D=\D+\E,
\label{eqn:relDE}\\
&\la W\vert(\alpha \E-\gamma \D)=\la W\vert ,
\label{eqn:relW1}\\
&(\beta \D-\delta \E)\vert V\ra =\vert V\ra.
\label{eqn:relV1}
\end{align}
The normalization constant $Z_L$ in (\ref{eqn:MPA}) is given by
\begin{align}
\label{eqn:defZ_L}
Z_L=\la W\vert \C^L\vert V\ra  
\end{align}
where $\C=\D+\E$.
We call $Z_L$ the partition function and $\C$ the transfer matrix. 
The reasons are as follows. In the similar way to the standard statistical mechanics, the bulk quantities of the system can be calculated from $Z_L$. More precisely, if we introduce the fugacity $\lambda$ and define $Z_L(\lambda)=\la W\vert (\lambda\D+\E)^L\vert V\ra$, the density $\rho$ and the density fluctuation $\Delta\rho^2$ are obtained from 
\begin{align}
&\rho = 
\frac{1}{L}\lambda\frac{\partial}{\partial\lambda}
\log Z_L(\lambda)\bigg\vert_{\lambda=1},
\label{def:rho}\\
&\Delta\rho^2 = 
\left(\frac{1}{L}\lambda\frac{\partial}{\partial\lambda}\right)^2
\log Z_L(\lambda)\bigg\vert_{\lambda=1}.
\label{def:rho^2}
\end{align}
The definition of the matrix $\C$ implies that it has all information on the possibility at each site, occupation and emptiness of a particle. Since it is also site-independent, the partition function is calculated from the eigenvectors of $\C$. Furthermore, the partition function is dominated by the contribution of the largest eigenvalue of $\C$ in the thermodynamic limit $L\to\infty$. Interestingly, phase transitions occur when the set of eigenvalues of $\C$ varies qualitatively. 
We shall come back to this point later. 

Among other physical quantities we are interested in the $n$-point functions. For example, the one-point function and the two-point function are respectively written as 
\begin{align}
\label{def:1Pfn}
& \la \tau_i\ra =
\frac{1}{Z_L}\la W\vert \C^{i-1}\D\C^{L-i}\vert V\ra,\\
\label{def:2Pfn}
& \la \tau_i\tau_j\ra =
\frac{1}{Z_L}\la W\vert \C^{i-1}\D\C^{j-i-1}\D\C^{L-j}\vert V\ra.
\end{align}
Here the bracket in the LHS means the average over the steady state distribution (\ref{eqn:MPA}). An important nonequilibrium quantity, the particle current $J$, is defined by the two-point function: $J=\la\tau_i(1-\tau_{i+1})-q(1-\tau_i)\tau_{i+1}\ra$. In terms of the matrix product expression, we have 
\begin{align}
J=\frac{Z_{L-1}}{Z_L}.
\end{align}
Indeed, the current $J$ takes nonzero value even in the steady state. Therefore, the ASEP is considered to be a nonequilibrium model.

\section{Known results}

We briefly review known results for the steady state of the ASEP with open boundaries. Most of them are found in \cite{USW04}. 
There, the Askey-Wilson polynomial plays a chief role. 
The Askey-Wilson polynomial $P_n(x)=P_n(x;a,b,c,d\vert q), n\in\Z_+$ is a $q$-orthogonal polynomial with four parameters $a, b, c, d$ besides $q$. It is regarded as the most general one in the hierarchy of the one-variable $q$-orthogonal polynomial family in the Askey scheme \cite{AW,GR,Koekoek}. 
The orthogonality relation for $P_n(x)$ is 
\begin{align}
\label{eqn:ortho}
\frac{1}{2\pi}\int_{-1}^1 \frac{w(x)}{\sqrt{1-x^2}} P_m(x)P_n(x)dx 
+\sum_{f=a,b,c,d}\sum_{k:1<\vert fq^k\vert\le \vert f\vert} 
w_k^{f} P_m(x_k^{f})P_n(x_k^{f})
=h_n\delta_{mn}
\end{align}
where, by using the notation of the $q$-shifted factorial $(a_1,\cdots,a_s;q)_n=\prod_{r=1}^s \prod_{k=0}^{n-1} (1-a_rq^k)$, 
\begin{gather}
w(\cos\theta)=
\frac{(e^{2i\theta},e^{-2i\theta};q)_\infty}
{(ae^{i\theta},ae^{-i\theta},be^{i\theta},be^{-i\theta},
ce^{i\theta},ce^{-i\theta},de^{i\theta},de^{-i\theta};q)_\infty} ,\\
\frac{h_n}{h_0}=
\frac{(1-q^{n-1}abcd)(q,ab,ac,ad,bc,bd,cd;q)_n}
{(1-q^{2n-1}abcd)(abcd;q)_n} ,\\
h_0=
\frac{(abcd;q)_\infty}{(q,ab,ac,ad,bc,bd,cd;q)_\infty} ,\\
x_k^{f}=
\frac{fq^k+(fq^k)^{-1}}{2} ,\\
w_k^{a}=
\frac{(a^{-2};q)_\infty}{(q,ab,ac,ad,a^{-1}b,a^{-1}c,a^{-1}d;q)_\infty}
\frac{(1-a^2q^{2k})(a^2,ab,ac,ad;q)_k}{(1-a^2)(q,ab^{-1}q,ac^{-1}q,ad^{-1}q;q)_\infty}
\(\frac{q}{abcd}\)^k ,
\end{gather}
and similarly for $f=b,c,d$. 
In the theory of the orthogonal polynomial, the three-term recurrence relation and the orthogonality relation are fundamental objects. The three-term recurrence relation is represented by an infinite-dimensional matrix called the Jacobi operator $\T$ as an eigenvalue equation, 
\begin{align}
\T\vert P(x)\ra =x\vert P(x)\ra 
\end{align}
where $\vert P(x)\ra = {}^t (P_0(x),P_1(x),\cdots)$. 
The fact is known as the spectral theorem that the spectrum of the Jacobi operator corresponds to the support of the orthogonal measure of the orthogonal polynomial (see for example \cite{Deift}). 
From (\ref{eqn:ortho}) we can read off the spectrum of $\T$ as $[-1,1]\cup\{x_k^f,f=a,b,c,d\big\vert 1<\vert fq^k\vert\le \vert f\vert, k\in\Z_+ \}$ for the Askey-Wilson polynomial. 

For the steady state of the ASEP, we found in \cite{USW04} that the transfer matrix $\C$ is directly related to the Jacobi operator of the Askey-Wilson polynomial $\T$: 
\begin{align}
\C=2+2\T
\end{align}
with parameters being set 
\begin{align}
\label{abcd}
a=\kappa_{\alpha,\gamma}^+,\ b=\kappa_{\beta,\delta}^+,\ 
c=\kappa_{\alpha,\gamma}^-,\ d=\kappa_{\beta,\delta}^-,
\end{align}
where 
\begin{align}
&\kappa_{\alpha,\gamma}^\pm =
\frac{1}{2\alpha}
\left[ (1-q-\alpha+\gamma)\pm
\sqrt{(1-q-\alpha+\gamma)^2+4\alpha\gamma}\right] ,\\
&\kappa_{\beta,\delta}^\pm =
\frac{1}{2\beta}
\left[ (1-q-\beta+\delta)\pm
\sqrt{(1-q-\beta+\delta)^2+4\beta\delta}\right] .
\end{align}
Therefore, the eigenvector $\vert x\ra$ of $\C$ is written in terms of the Askey-Wilson polynomials and the spectrum of $\C$ is determined as $[0,4]\cup\{2+2x_k^f,f=a,b\big\vert 1<\vert fq^k\vert\le \vert f\vert, k\in\Z_+ \}$, noticing that $\vert c\vert,\vert d\vert <1$. 
In fact the partition function (\ref{eqn:defZ_L}) is calculated to be in the integral form, 
\begin{align}
Z_L=
\oint_C\frac{dz}{4\pi iz}
\frac{(z^2,z^{-2};q)_\infty
[(1+z)(1+z^{-1})/(1-q)]^L}
{(az,a/z,bz,b/z,cz,c/z,dz,d/z;q)_\infty} 
\label{eqn:Z_L}
\end{align}
where the integral contour $C$ is a closed path which 
encloses the poles at $z=aq^k$, $bq^k$, $cq^k$, $dq^k$ $(k\in \Z_+)$ 
and excludes the poles at $z=(aq^k)^{-1}$, $(bq^k)^{-1}$, $(cq^k)^{-1}$, $(dq^k)^{-1}$ $(k\in \Z_+)$. This integral is actually the moment integral with respect to the weight function of the Askey-Wilson polynomial. 

Although the integral of (\ref{eqn:Z_L}) is difficult to evaluate, physically important thermodynamic limit $L\to\infty$ makes it possible to obtain the leading contribution. In this limit, the largest eigenvalue of $\C$ dominates the contribution to the partition function. Indeed, it is clear if we expand 
\begin{align*}
\la W\vert \C^L\vert V\ra &=\sum_x \la W\vert \C^L\vert x\ra\la x\vert V\ra \\
&=\sum_x \la W\vert x\ra x^L\la x\vert V\ra \\
&\simeq \mathrm{const.} x_{\mathrm{max}}^L .
\end{align*}
Taking account of the largest eigenvalue, we find that there are three phases.
The bulk quantities for each phase are eventually obtained as the following. \\
(A) $a>1$ and $a>b$: $x_{\mathrm{max}}=(1+a)(1+a^{-1})$, 
\begin{gather}
\label{eqn:Z_L-A}
Z_L\simeq 
\frac{(a^{-2};q)_\infty}
{(q,ab,ac,ad,a^{-1}b,a^{-1}c,a^{-1}d;q)_\infty}
\left[\frac{(1+a)(1+a^{-1})}{1-q}\right]^L ,\\
J\simeq 
(1-q)\frac{a}{(1+a)^2},\qquad
\rho \simeq 
\frac{1}{1+a},\qquad
\Delta\rho^2 \simeq 
\frac{a}{(1+a)^2L}.
\end{gather}
(B) $b>1$ and $b>a$: $x_{\mathrm{max}}=(1+b)(1+b^{-1})$, 
\begin{gather}
\label{eqn:Z_L-B}
Z_L\simeq 
\frac{(b^{-2};q)_\infty}
{(q,ba,bc,bd,b^{-1}a,b^{-1}c,b^{-1}d;q)_\infty}
\left[\frac{(1+b)(1+b^{-1})}{1-q}\right]^L ,\\
J\simeq (1-q)\frac{b}{(1+b)^2},\qquad
\rho\simeq 
\frac{b}{1+b},\qquad
\Delta\rho^2 \simeq 
\frac{b}{(1+b)^2L}.
\end{gather}
(C) $a<1$ and $b<1$: $x_{\mathrm{max}}=4$, 
\begin{gather}
\label{eqn:Z_L-C}
Z_L \simeq 
\frac{(q;q)_\infty^2}{(a,b,c,d;q)_\infty^2}
\frac{4}{\sqrt{\pi}L^{3/2}}
\left(\frac{4}{1-q}\right)^L ,\\
J\simeq 
\frac{1-q}{4},\qquad
\rho \simeq 
\frac{1}{2},\qquad
\Delta\rho^2 \simeq 
\frac{1}{8L}.
\end{gather}
Note that a phase transition occurs when the largest eigenvalue $x_{\mathrm{max}}$ of the transfer matrix $\C$ changes from one to the other. 

Now, let us consider the correlation functions. We have the following integral formula for the $n$-point function, 
\begin{eqnarray} \label{eqn:npoint}
&& \langle \tau_{j_1}\cdots\tau_{j_n}\rangle=
\frac{1}{(1-q)^LZ_L}
\lim_{\xi_1,\cdots,\xi_n\to1} 
\left[ \prod_{m=1}^{n+1} \oint_{C_m} \frac{dz_m}{4\pi iz_m} \right] \nonumber\\
&&\qquad\times
	\frac{\displaystyle{\prod_{m=1}^{n+1}(z_m^2,z_m^{-2};q)_\infty
	[(1+z_m)(1+1/z_m)]^{j_m-j_{m-1}-1}}}
	{(az_1,a/z_1,cz_1,c/z_1,
	bz_{n+1},b/z_{n+1},dz_{n+1},d/z_{n+1};q)_\infty} 
\nonumber\\
&&\qquad\times
	\prod_{m=1}^{n}
	\bigg[
	\frac{1}{(z_m^2,z_m^{-2};q)_\infty} \delta(z_{m+1}-z_m)
\nonumber\\
&&\qquad\qquad+
	\frac{(q;q)_\infty^2 (z_{m+1}+1/z_{m+1}-\xi_m z_m-\xi_m /z_m)}
	{(\xi_m z_mz_{m+1},\xi_m z_m/z_{m+1},
	\xi_m z_{m+1}/z_m,\xi_m /z_mz_{m+1};q)_\infty}\bigg]
\end{eqnarray}
where $j_0=0,j_{n+1}=L+1$ in the product. 
Regularization parameters $\xi_m$'s are included since otherwise this integral 
becomes a singular integral. 
The contour $C_m(m=2,\cdots,n+1)$ encloses the poles at 
$z_m=\xi_{m-1} z_{m-1}q^k,$ $\xi_{m-1} z_{m-1}^{-1}q^k$ $(k\in\Z_+)$ 
and excludes the poles at 
$z_m=(\xi_{m-1} z_{m-1}q^k)^{-1},$ $(\xi_{m-1} z_{m-1}^{-1}q^k)^{-1}$ $(k\in\Z_+)$. 
In addition, for $m=1$, $C_1$ encloses the poles at 
$z_1=aq^k,cq^k$ $(k\in\Z_+)$ and excludes the poles at 
$z_1=(aq^k)^{-1},(cq^k)^{-1}$ $(k\in\Z_+)$. 
Also, $C_{n+1}$ encloses the poles at $z_{n+1}=bq^k,dq^k$ $(k\in\Z_+)$ 
and excludes the poles at $z_{n+1}=(bq^k)^{-1},(dq^k)^{-1}$ $(k\in\Z_+)$. 
In the thermodynamic limit, 
the second largest eigenvalue also comes into the game. 
For the two-point function (\ref{def:2Pfn}), we have an expansion 
\begin{align*}
&\langle W\vert \C^{i-1}\D\C^{j-i-1}\D
\C^{L-j}\vert V\rangle\\
&\simeq
x_{\mathrm{max}}^{L-2} \la W\vert x_{\mathrm{max}}\ra 
\la x_{\mathrm{max}} \vert D\vert x_{\mathrm{max}}\ra 
\la x_{\mathrm{max}} \vert D\vert x_{\mathrm{max}}\ra 
\la x_{\mathrm{max}} \vert V\ra\\
&\quad +x_{\mathrm{max}}^{L-2} 
\left(\frac{x_{\mathrm{2nd}}}{x_{\mathrm{max}}}\right)^{i-1}
\la W\vert x_{\mathrm{2nd}}\ra 
\la x_{\mathrm{2nd}} \vert D\vert x_{\mathrm{max}}\ra 
\la x_{\mathrm{max}} \vert D\vert x_{\mathrm{max}}\ra 
\la x_{\mathrm{max}} \vert V\ra\\
&\quad +x_{\mathrm{max}}^{L-2} 
\left(\frac{x_{\mathrm{2nd}}}{x_{\mathrm{max}}}\right)^{j-i-1}
\la W\vert x_{\mathrm{max}}\ra 
\la x_{\mathrm{max}} \vert D\vert x_{\mathrm{2nd}}\ra 
\la x_{\mathrm{2nd}} \vert D\vert x_{\mathrm{max}}\ra 
\la x_{\mathrm{max}} \vert V\ra\\
&\quad +x_{\mathrm{max}}^{L-2} 
\left(\frac{x_{\mathrm{2nd}}}{x_{\mathrm{max}}}\right)^{L-j}
\la W\vert x_{\mathrm{max}}\ra 
\la x_{\mathrm{max}} \vert D\vert x_{\mathrm{max}}\ra 
\la x_{\mathrm{max}} \vert D\vert x_{\mathrm{2nd}}\ra 
\la x_{\mathrm{2nd}} \vert V\ra.
\end{align*}
From this expansion, it is found that the ratio $x_{\mathrm{2nd}}/x_{\mathrm{max}}$ determines the decay scale of the correlation. 
However, the second largest eigenvalue is defined only for the phases (A) and (B) since the spectrum of $\C$ is continuous for the phase (C). 
For this reason we investigate in detail the correlation functions in the thermodynamic limit in the subsequent section.

\section{Correlation functions}

We will study the $n$-point function. 
In the following, suppose that the site numbers are of order or suborder $L$, and take a thermodynamic limit $L\to\infty$. 

First, the phases (A) and (B). 
In these cases, the second largest eigenvalue of $\C$ is separated from the largest one and can be explicitly read off from the spectrum. 
It is not necessary to touch the integral (\ref{eqn:npoint}) directly. 
The correlation functions decay exponentially and the correlation length is determined by $\xi=(\ln x_{\mathrm{max}}/x_{\mathrm{2nd}})^{-1}$. 
The phase (A) is further classified into three sub-phases. 
\\
\noindent
(A$_1$) $aq>1,\ aq>b$,
\begin{align}
x_{\mathrm{max}}=(1+a)(1+a^{-1}),\qquad
x_{\mathrm{2nd}}=(1+aq)(1+(aq)^{-1}).
\end{align}
(A$_2$) $a>b>aq,\ b>1$,
\begin{align}
x_{\mathrm{max}}=(1+a)(1+a^{-1}),\qquad
x_{\mathrm{2nd}}=(1+b)(1+b^{-1}).
\end{align}
(A$_3$) $a>1>aq,\ b<1$,
\begin{align}
x_{\mathrm{max}}=(1+a)(1+a^{-1}),\qquad
x_{\mathrm{2nd}}=4.
\end{align}
Because of the particle-hole symmetry, the phase (B) is classified in the same way except $a\leftrightarrow b$ in the above. 

\begin{figure}[htb]
\begin{center}
\includegraphics[scale=0.5]{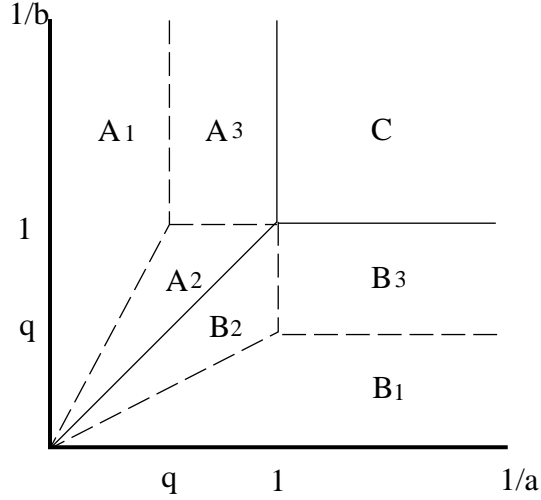}
\caption{Phase diagram for the correlation length. 
The solid lines distinguish the three phases (A), (B) and (C). 
The dashed lines separate the three sub-phases in (A) and (B).}
\label{fig:phase}
\end{center}
\end{figure}

Next, the phase (C). 
As mentioned before, since the spectrum of the transfer matrix $\C$ is continuous, the second largest eigenvalue is not defined. Thus, we should take care of the integral representation. The behavior of the correlation function turns out to be something different from the exponential decay. We can identify the correlation length with infinity in this case, and finally we obtain a phase diagram with respect to the correlation length as Figure \ref{fig:phase}. 
The horizontal (resp. vertical) axis corresponds to the left (resp. right) boundary. 
As $1/a$ increases, the input from the left boundary tends to increase; as $1/b$ increases, the output into the right boundary tends to increase. 
The special case $\gamma=\delta=0$ was argued and a similar phase diagram was obtained in \cite{SS99}. 

We consider the finite-size correction of the correlation functions. 
Taking a close look at the integral (\ref{eqn:npoint}), we can verify a decomposition 
\begin{align}
\langle \tau_{j_1}\cdots\tau_{j_n}\rangle=
\sum_{k=0}^n \sum_{\{i_1,\cdots,i_k\}\subset\{j_1,\cdots,j_n\}} 
\rho^{n-k} \la\delta\tau_{i_1}\cdots\delta\tau_{i_k}\ra .
\end{align}
By definition, $\la\delta\tau_{i_1}\cdots\delta\tau_{i_k}\ra$ is the term represented by some integral which depends only on $\{i_1,\cdots,i_k\}$ out of $\{j_1,\cdots,j_n\}$ and is independent of $n$. 
These terms come from the existence of the boundaries. 
Therefore, we can identify them as the finite-size corrections. 
The $n$-point correction term of the correlation functions is written in the following integral form, 
\begin{align} \label{eqn:npoint-C}
&\la\delta\tau_{j_1}\cdots\delta\tau_{j_n}\ra 
=\frac{(q;q)_\infty^{2n}}{(1-q)^LZ_L}
\notag\\
&\qquad\times
\left[ \prod_{m=1}^{n+1} \int_0^\pi \frac{d\theta_m}{2\pi}\right] 
\frac{\displaystyle{\prod_{m=1}^{n+1} (e^{2i\theta_m},e^{-2i\theta_m};q)_\infty}%
(2+2\cos\theta_m)^{j_m-j_{m-1}-1}}
{(ae^{i\theta_1},ae^{-i\theta_1},ce^{i\theta_1},ce^{-i\theta_1},
be^{i\theta_1},be^{-i\theta_1},de^{i\theta_1},de^{-i\theta_1};q)_\infty}
\notag \\
&\qquad\times
\prod_{m=1}^n 
\frac{(2\cos\theta_{j_{m+1}}-2\cos\theta_{j_m})}
{(e^{i(\theta_{j_m}+\theta_{j_{m+1}})},
e^{i(\theta_{j_m}-\theta_{j_{m+1}})},
e^{-i(\theta_{j_m}-\theta_{j_{m+1}})},
e^{-i(\theta_{j_m}+\theta_{j_{m+1}})};q)_\infty}
\end{align}
where $j_0=0,j_{n+1}=L+1$ in the product. 
In the thermodynamic limit, the contribution to this integral is concentrated on $\theta_m=0, m=1,\cdots,n+1$ due to the presence of $(2+2\cos\theta_m)^{j_m-j_{m-1}-1}$. Approximating (\ref{eqn:npoint-C}) around this point by the saddle point method and substituting the result (\ref{eqn:Z_L-C}), we arrive at a simpler expression, 
\begin{align}
\label{eqn:integral1}
\la\delta\tau_{j_1}\cdots\delta\tau_{j_n}\ra 
\simeq 
L^{3/2}\frac{\sqrt{\pi}}{(2\pi)^{n+1}}
\prod_{m=1}^{n+1} \int_0^\infty d\theta_m 
\frac{\displaystyle{\prod_{m=1}^{n+1} \theta_m^2 \exp\left[-\frac{1}{4}(j_m-j_{m-1}-1)\theta_m^2\right]}}
{\displaystyle{\prod_{m=1}^n(\theta_m^2-\theta_{m+1}^2)}} .
\end{align}
The integrands have singularities at $\theta_m=\theta_{m+1}$, but this multifold integral should have a certain finite value because we are evaluating a finite-valued physical quantity. 
To clarify this point, we rewrite the integral in the RHS; in fact it is equivalent to another form, 
\begin{align}
\label{eqn:integral2}
&\mathcal{I}_n(\xi_1,\cdots,\xi_{n+1})
\notag \\
&\ =\ -\(\frac{\sqrt\pi}{2}\)^{n+1}
\prod_{m=1}^{n+1}\left(\frac{\partial}{\partial \xi_m}\right)
\prod_{m=1}^n \left[ \int_0^{\sum_{k=1}^m \xi_k-\sum_{k=1}^{m-1} \eta_k}
\frac{d\eta_m}{\sqrt{\eta_m}}\right]
\frac{1}{\sqrt{\sum_{k=1}^{n+1} \xi_k-\sum_{k=1}^{n} \eta_k}}
\end{align}
with $\xi_m=(j_m-j_{m-1}-1)/4,\ m=1,\cdots,n+1$. 
In turn, the integral in (\ref{eqn:integral2}) is integrable and definitely has a finite value. 
Rewriting of the integral is performed as follows. 
First of all, we have the Gaussian integral
\begin{align*}
\int_0^\infty e^{-\eta(\theta^2-\varphi^2)} d\theta = 
\frac{\sqrt\pi}{2\sqrt\eta}e^{\eta\varphi^2}. 
\end{align*}
Then, integrating the above over $\eta$ from $0$ to $\xi$ yields 
\begin{align*}
-\int_0^\infty \frac{e^{-\xi(\theta^2-\varphi^2)}}{\theta^2-\varphi^2} d\theta
+\int_0^\infty \frac{d\theta}{\theta^2-\varphi^2}
=\int_0^\xi \frac{\sqrt\pi}{2\sqrt\eta}e^{\eta\varphi^2}d\eta . 
\end{align*}
The second term of the LHS is equal to $\frac{\pi^2}{2}\delta (\varphi )$. Now multiplying $e^{-\xi\varphi^2}$ on both sides and then differentiating by $\xi$, we have 
\begin{align*}
\int_0^\infty \frac{\theta^2 }{\theta^2-\varphi^2}\ e^{-\xi\theta^2} d\theta
= \frac{\partial}{\partial\xi}
\left[ e^{-\xi\varphi^2}\int_0^\xi \frac{\sqrt\pi}{2\sqrt\eta}e^{\eta\varphi^2}d\eta\right] .
\end{align*}
Note that we can drop $\varphi^2 e^{-\xi\varphi^2}\delta(\varphi)$. 
Applying this formula to the integral of $\theta_1$, then $\theta_2$, $\cdots$, $\theta_n$, we get
\begin{align*}
&\prod_{m=1}^{n+1} \int_0^\infty d\theta_m 
\frac{\displaystyle{\prod_{m=1}^{n+1} \theta_m^2 
\exp\left[-\xi_m\theta_m^2\right]}}
{\displaystyle{\prod_{m=1}^n(\theta_m^2-\theta_{m+1}^2)}} \\
&=
\(\frac{\sqrt\pi}{2}\)^{n}
\prod_{m=1}^{n}\left(\frac{\partial}{\partial \xi_m}\right)
\prod_{m=1}^n \left[ \int_0^{\sum_{k=1}^m \xi_k-\sum_{k=1}^{m-1} \eta_k}
\frac{d\eta_m}{\sqrt{\eta_m}}\right] \\
&\qquad\times
\int_0^\infty \theta_{n+1}^2 \exp\left[-\(\sum_{k=1}^{n+1} \xi_k-\sum_{k=1}^{n} \eta_k\)\theta_{n+1}^2\right] d\theta_{n+1}. 
\end{align*}
Finally, performing the $\theta_{n+1}$ integral, we obtain (\ref{eqn:integral2}). 
After all, we have the following result. 
\begin{align}
\la\delta\tau_{j_1}\cdots\delta\tau_{j_n}\ra 
\simeq 
L^{3/2}\frac{\sqrt{\pi}}{(2\pi)^{n+1}}
\mathcal{I}_n\(
\frac{j_1-1}{4},\frac{j_2-j_1-1}{4},\cdots,\frac{j_n-j_{n-1}-1}{4},\frac{L-j_n}{4}\).
\end{align}
From this expression, it is found that the correlation functions are independent of the boundary parameters $a,b,c,d$ and $q$. Moreover, one can show that the $n$-point correction $\la\delta\tau_{j_1}\cdots\delta\tau_{j_n}\ra $ is of order $L^{-n/2}$ if $j_1,\cdots,j_n$ are of order $L$. 

We obtain exact expressions for the $n$-point corrections by calculating (\ref{eqn:integral2}) explicitly. We show the first few examples: 
\begin{gather}
\la \delta\tau_{i}\ra \simeq
-\frac{1}{L^{1/2}\sqrt{\pi}}\frac{i-L/2}{\sqrt{i(L-i)}} ,\\
\la \delta\tau_{i}\delta\tau_{j}\ra \simeq
-\frac{1}{2\pi} \Bigg[ L^{-1/2} 
\frac{\sqrt{j-i}(-i^2+(j-i)^2+i(L-j)-(L-j)^2)}
{2\sqrt{i(L-j)}j(L-i)} \notag\\
\qquad\qquad\qquad\qquad
+L^{-1} \frac{3}{2} \arctan \sqrt{\frac{i(L-j)}{(j-i)L}}\Bigg] ,
\end{gather}
\begin{gather}
\la \delta\tau_{i}\delta\tau_{j}\delta\tau_{k}\ra \simeq
-\frac{1}{2^2\pi^{3/2}}\Bigg[
L^{-1/2}
\frac{\sqrt{(j-i)(k-j)}}{2\sqrt{i(L-k)}j(L-j)k(L-i)}\notag\\
\times
\big\{
i^3+i^2(j-i)-i(j-i)^2-(j-i)^3-i(j-i)(k-j)-(j-i)^2(k-j)\notag\\
+(j-i)(k-j)^2+(k-j)^3-i^2(L-k)-i(j-i)(L-k)+i(k-j)(L-k)\notag\\
+(j-i)(k-j)(L-k)+(k-j)^2(L-k)+i(L-k)^2-(k-j)(L-k)-(L-k)^3
\big\}\notag\\
+L^{-3/2}\frac{(L-k)^2+6k(L-k)-3k^2}{2\sqrt{L-k}k^{3/2}}
\arctan\sqrt{\frac{i(k-j)}{(j-i)k}}\notag\\
-L^{-3/2}\frac{i^2+6i(L-i)-3(L-i)^2}{2\sqrt{i}(L-i)^{3/2}}
\arctan\sqrt{\frac{(j-i)(L-k)}{(k-j)(L-i)}}
\Bigg] .
\end{gather}

\begin{figure}
\begin{center}
\includegraphics[scale=1.2]{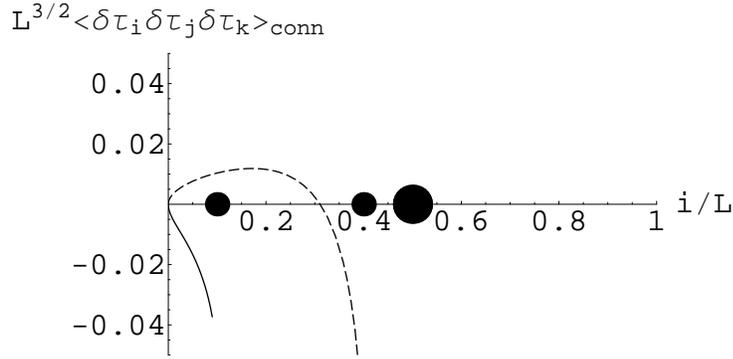}
\caption{
The connected part $\la \delta\tau_{i}\delta\tau_{j}\delta\tau_{k}\ra_{\mathrm{conn}}$ for large $L$. 
The two curves are plotted as a function of the position of the leftmost particle $i$ with the position of the middle particle $j$ and that of the rightmost one $k$ fixed at the place of the small black circles and the big black circle respectively. 
The solid line is for $j=0.1L$ and $k=0.5L$. The dashed line is for $j=0.4L$ and $k=0.5L$. 
}
\label{fig:c3point}
\end{center}
\end{figure}

Physically, fluctuation of the correlation function is related to 
the connected part, 
\begin{align}
\la \delta\tau_{j_1}\cdots\delta\tau_{j_n}\ra _{\mathrm{conn}}
=\big\la \(\delta\tau_{j_1}-\la\delta\tau_{j_1}\ra\) \cdots 
\(\delta\tau_{j_1}-\la\delta\tau_{j_1}\ra\) \big\ra . 
\end{align}
Figure \ref{fig:c3point} illustrates the behavior of the 
connected three-point correction term $\la \delta\tau_{i}\delta\tau_{j}\delta\tau_{k}\ra_{\mathrm{conn}}$ for large $L$. 
It is interesting that there can be a maximal or minimal point for the particle occupation according to the positions of the other particles. 
One also notices that $\la \delta\tau_{i}\delta\tau_{j}\delta\tau_{k}\ra_{\mathrm{conn}}$ vanishes at the boundary. This is a physical consequence of the fact that particles are in equilibrium with the reservoir at the boundary. 
We can show in general that the connected $n$-point functions vanish at the boundaries. 
From (\ref{eqn:integral2}), we directly have 
\begin{align*}
&\mathcal{I}_n(\xi_1,\cdots,\xi_{n+1})= \notag\\
&\frac{\sqrt\pi}{2}\left[
\frac{1}{\sqrt{\xi_1}}\mathcal{I}_{n-1}(\xi_2,\cdots,\xi_{n+1})
+\int_0^{\xi_1} \frac{d\eta}{\sqrt{\eta}}\frac{\partial}{\partial \xi_1}
\mathcal{I}_{n-1}(\xi_1+\xi_2-\eta,\cdots,\xi_{n+1})
\right] .
\end{align*}
Note if we take $\xi_1\to0$, only the first term of RHS is singular and the second term vanishes. Then, we have 
$\lim_{i\to0}
\la \delta\tau_{i}\delta\tau_{j}\cdots\delta\tau_{k}\ra
-\la \delta\tau_{i}\ra \la \delta\tau_{j}\cdots\delta\tau_{k}\ra =0$ 
because 
$\la \delta\tau_i\ra \sim \frac{1}{2\sqrt{\pi i}},\ i\to0$, 
and furthermore 
$\lim_{i\to0}
\Big\la \delta\tau_{i}(\delta\tau_{j}-\la\delta\tau_{j}\ra)
\cdots(\delta\tau_{k}-\la\delta\tau_{k}\ra)\Big\ra
-\la \delta\tau_{i}\ra \Big\la (\delta\tau_{j}-\la\delta\tau_{j}\ra)
\cdots(\delta\tau_{k}-\la\delta\tau_{k}\ra)\Big\ra =0$ ,
which establishes 
\begin{align}
\lim_{j_1\to0}
\la \delta\tau_{j_1}\cdots\delta\tau_{j_n}\ra _{\mathrm{conn}}=0.
\end{align}
Similarly, $\la \delta\tau_{j_1}\cdots\delta\tau_{j_n}\ra_{\mathrm{conn}}$ vanishes also in the limit $j_n\to L$. 

What we have calculated here are the leading terms of the finite-size corrections of $\la \delta\tau_{j_1}\cdots$ $\delta\tau_{j_n}\ra _{\mathrm{conn}}$. 
It is also possible to have $1/L$-expansion of it by elaborating approximations of (\ref{eqn:npoint-C}) and (\ref{eqn:Z_L-C}).

\section{Conclusion}

We have investigated the correlation functions for the steady state of the ASEP with open boundaries in the thermodynamic limit. 
Main results are summarized as follows: 
1) a phase diagram is given with respect to the correlation length $\xi$, 
2) for the case $\xi=\infty$, the leading terms of the finite-size corrections are calculated accurately. 

For the future problem, we are interested in the time-correlations of the ASEP. In order to calculate the time-correlations, one should know all the eigenvalues and eigenvectors of the time evolution operator $H$. 
At present, we have an exact information only for the steady state. 
It would be possible to solve the eigenvalue problem for $H$ 
by means of the Bethe ansatz method and evaluate the time-correlation functions for the ASEP with open boundaries. 

In the nonequilibrium statistical mechanics, we like to have the general theory for the relations among nonequilibrium quantities. 
For the purpose, the study of the current fluctuation is primarily important. 
An attempt to calculate the current fluctuations of the ASEP with open boundaries was made for the symmetric case $q=1$ \cite{DDR}. 
There, the authors considered a hierarchy of equations for the correlations without directly solving the eigenvalue problem for $H$. 
Whether their approach is applicable to the asymmetric case $q\neq 1$ is one of open questions.

\label{lastpage}

\end{document}